\input harvmac.tex
\input epsf.tex


\def\figin{\epsfcheck\figin}\def\figins{\epsfcheck\figins}
\def\epsfcheck{\ifx\epsfbox\UnDeFiNeD
\message{(NO epsf.tex, FIGURES WILL BE IGNORED)}
\gdef\figin##1{\vskip0in}\gdef\figins##1{\hskip.0in}
\else\message{(FIGURES WILL BE INCLUDED)}%
\gdef\figin##1{##1}\gdef\figins##1{##1}\fi}
\def\DefWarn#1{}
\def\figinsert{\goodbreak\midinsert}
\def\ifig#1#2#3{\DefWarn#1\xdef#1{fig.~\the\figno}
\writedef{#1\leftbracket fig.\noexpand~\the\figno}%
\figinsert\figin{\centerline{#3}}\medskip\centerline{\vbox{\baselineskip12pt
\advance\hsize by -1truein\noindent\footnotefont{\bf
Fig.~\the\figno:} #2}}
\bigskip\endinsert\global\advance\figno by1}



\def\frac#1#2{{#1 \over #2}}
\def\text#1{#1}
\def\half{{1\over 2}}
\def\Id{{1\!\!1}}


\lref\nostring{
  I.~Bena, E.~Gorbatov, S.~Hellerman, N.~Seiberg and D.~Shih,
  JHEP {\bf 0611}, 088 (2006)
  [arXiv:hep-th/0608157].
}

\lref\nostringa{
  S.~Murthy,
  arXiv:hep-th/0703237.
}

\lref\clastring{ S.~Franco, I.~Garcia-Etxebarria and A.~M.~Uranga,
  JHEP {\bf 0701}, 085 (2007)
  [arXiv:hep-th/0607218].
  }

\lref\clastringa{ H.~Ooguri and Y.~Ookouchi,
  Phys.\ Lett.\  B {\bf 641}, 323 (2006)
  [arXiv:hep-th/0607183].
}

\lref\HoweEU{
  P.~S.~Howe, N.~D.~Lambert and P.~C.~West,
  Phys.\ Lett.\  B {\bf 418}, 85 (1998)
  [arXiv:hep-th/9710034].
}

\lref\WittenSC{
  E.~Witten,
  Nucl.\ Phys.\  B {\bf 500}, 3 (1997)
  [arXiv:hep-th/9703166].
}

\lref\AhnYQ{
  C.~Ahn,
  Class.\ Quant.\ Grav.\  {\bf 24}, 1359 (2007)
  [arXiv:hep-th/0608160];
  Phys.\ Lett.\  B {\bf 647}, 493 (2007)
  [arXiv:hep-th/0610025];
  Class.\ Quant.\ Grav.\  {\bf 24}, 3603 (2007)
  [arXiv:hep-th/0702038];
  arXiv:hep-th/0703015;
  arXiv:0704.0121 [hep-th];
  arXiv:0705. 0056 [hep-th];
  arXiv:0706.0042 [hep-th];
  arXiv:0707.0092 [hep-th].
}

\lref\twobe{
  R.~Argurio, M.~Bertolini, S.~Franco and S.~Kachru,
  JHEP {\bf 0701}, 083 (2007)
  [arXiv:hep-th/0610212];
  JHEP {\bf 0706}, 017 (2007)
  [arXiv:hep-th/0703236].
}

\lref\twobea{
   T.~Kawano, H.~Ooguri and Y.~Ookouchi,
  arXiv:0704.1085 [hep-th].
}

\lref\twobeb{
    M.~Serone and A.~Westphal,
  arXiv:0707.0497 [hep-th].
}

\lref\nogauge{
  M.~Aganagic, C.~Beem, J.~Seo and C.~Vafa,
  arXiv:hep-th/0610249.
    J.~J.~Heckman, J.~Seo and C.~Vafa,
  arXiv:hep-th/0702077.
    J.~J.~Heckman and C.~Vafa,
  arXiv:0707.4011 [hep-th].
  }

\lref\beem{
    M.~Aganagic, C.~Beem and B.~Freivogel,
  arXiv:0708.0596 [hep-th].
}

\lref\nogaugea{
    R.~Tatar and B.~Wetenhall,
  JHEP {\bf 0702}, 020 (2007)
  [arXiv:hep-th/0611303];
  arXiv: 0707.2712 [hep-th].
}

\lref\nogaugeb{
  A.~Giveon and D.~Kutasov,
  Nucl.\ Phys.\  B {\bf 778}, 129 (2007)
  [arXiv:hep-th/0703135].
}

\lref\nogaugec{
  M.~R.~Douglas, J.~Shelton and G.~Torroba,
  arXiv:0704.4001 [hep-th].
}

\lref\OoguriIU{
  H.~Ooguri, Y.~Ookouchi and C.~S.~Park,
  arXiv:0704.3613 [hep-th].
}

\lref\PastrasQR{
  G.~Pastras,
  arXiv:0705.0505 [hep-th].

}

\lref\deBoerBY{
  J.~de Boer, K.~Hori, H.~Ooguri and Y.~Oz,
  Nucl.\ Phys.\  B {\bf 522}, 20 (1998)
  [arXiv:hep-th/9801060].
}

\lref\WittenEP{
  E.~Witten,
  Nucl.\ Phys.\  B {\bf 507}, 658 (1997)
  [arXiv:hep-th/9706109].
}

\lref\MarsanoFE{
  J.~Marsano, K.~Papadodimas and M.~Shigemori,
  arXiv:0705.0983 [hep-th].
}

\lref\CachazoZK{
  F.~Cachazo, N.~Seiberg and E.~Witten,
  JHEP {\bf 0302}, 042 (2003)
  [arXiv:hep-th/0301006].
}

\lref\JanikHK{
  R.~A.~Janik,
  Phys.\ Rev.\  D {\bf 69}, 085010 (2004)
  [arXiv:hep-th/0311093].
}

\lref\GiveonSR{
  A.~Giveon and D.~Kutasov,
  Rev.\ Mod.\ Phys.\  {\bf 71}, 983 (1999)
  [arXiv:hep-th/9802067].
}

\lref\deBoerBY{
  J.~de Boer, K.~Hori, H.~Ooguri and Y.~Oz,
  Nucl.\ Phys.\  B {\bf 522}, 20 (1998)
  [arXiv:hep-th/9801060].
}

\lref\IntriligatorPU{ K.~I.~Izawa and T.~Yanagida,
  Prog.\ Theor.\ Phys.\  {\bf 95}, 829 (1996)
  [arXiv:hep-th/9602180].
  K.~A.~Intriligator and S.~D.~Thomas,
  Nucl.\ Phys.\  B {\bf 473}, 121 (1996)
  [arXiv:hep-th/9603158].
}

\lref\ISS{
  K.~Intriligator, N.~Seiberg and D.~Shih,
  JHEP {\bf 0604}, 021 (2006)
  [arXiv:hep-th/0602239].
}

\lref\IntriligatorCP{
  K.~Intriligator and N.~Seiberg,
  arXiv:hep-ph/0702069.
}

\lref\deBoerZY{
  J.~de Boer, K.~Hori, H.~Ooguri and Y.~Oz,
  Nucl.\ Phys.\  B {\bf 518}, 173 (1998)
  [arXiv:hep-th/9711143].
}

\lref\HoweEU{
  P.~S.~Howe, N.~D.~Lambert and P.~C.~West,
  Phys.\ Lett.\  B {\bf 418}, 85 (1998)
  [arXiv:hep-th/9710034].
}

\lref\HoriAB{
  K.~Hori, H.~Ooguri and Y.~Oz,
  Adv.\ Theor.\ Math.\ Phys.\  {\bf 1}, 1 (1998)
  [arXiv:hep-th/9706082].

}

\lref\BonelliRY{
  G.~Bonelli, M.~Matone and M.~Tonin,
  Phys.\ Rev.\  D {\bf 55}, 6466 (1997)
  [arXiv:hep-th/9610026].
}

\lref\deBoerAP{
  J.~de Boer and Y.~Oz,
  Nucl.\ Phys.\  B {\bf 511}, 155 (1998)
  [arXiv:hep-th/9708044].
}

\lref\GiudiceNI{
  G.~F.~Giudice and R.~Rattazzi,
  Nucl.\ Phys.\  B {\bf 511}, 25 (1998)
  [arXiv:hep-ph/9706540].
}

\lref\AraiMD{
  M.~Arai, C.~Montonen, N.~Okada and S.~Sasaki,
  arXiv:0708.0668 [hep-th].
}


\Title{\vbox{}} {\vbox{\centerline{Supersymmetry Breaking Vacua
from }
\medskip
\centerline{M Theory Fivebranes} }}
\smallskip
\centerline{Luca Mazzucato, Yaron Oz and Shimon Yankielowicz}
\smallskip
\bigskip
\centerline{\it Raymond and Beverly Sackler Faculty of Exact
Sciences }
\smallskip
 \centerline{\it School of Physics and Astronomy}
\smallskip
 \centerline{\it Tel-Aviv University, Ramat-Aviv 69978, Israel}
\medskip

\bigskip
\vskip 1cm


\vskip 0.5cm

\noindent We consider intersecting brane configurations realizing
${\cal N}=2$ supersymmetric gauge theories broken to ${\cal N}=1$
by multitrace superpotentials, and softly to ${\cal N}=0$. We
analyze, in the framework of M5-brane wrapping a curve, the
supersymmetric vacua and the analogs of spontaneous supersymmetry
breaking and soft supersymmetry breaking in gauge theories. We
show that the M5-brane does not exhibit the analog of metastable
spontaneous supersymmetry breaking, and does not have
non-holomorphic minimal volume curves with holomorphic boundary
conditions. However, we find that any point in the ${\cal N}=2$
moduli space can be rotated to a non-holomorphic minimal volume
curve, whose boundary conditions break supersymmetry. We interpret
these as the analogs of soft supersymmetry breaking vacua in the
gauge theory.

 \vskip 0.5cm

\Date{September 2007}


\newsec{Introduction and Summary}

A promising candidate for new physics beyond the Standard Model is
supersymmetry, which offers a solution to the hierarchy problem, a
unification of gauge couplings and a dark matter candidate.
Supersymmetry is broken in nature and one of the most important
problems is to understand the mechanism that leads to this
breaking.

One way of breaking supersymmetry is by adding explicit soft
supersymmetry breaking terms to the supersymmetric Lagrangian, as
in the case of the MSSM. Others are spontaneous supersymmetry
breaking mechanisms, which are particularly interesting when
supersymmetry is broken dynamically. A new paradigm for dynamical
supersymmetry breaking has been advocated by ISS \ISS, in which
the theory contains both supersymmetric vacua and also vacua that
break supersymmetry dynamically. In these scenarios, the
supersymmetry breaking vacuum is meta-stable. It has recently
become clear that this phenomenon is rather generic in
supersymmetric gauge theories (for a review and a discussion of
recent developments see \IntriligatorCP).

An important question is whether and how these supersymmetry
breaking mechanisms can be realized in M/string theory.  One
framework to address this question is in the intersecting branes
setup (for a review see \GiveonSR). Here, one typically engineers
the gauge theory on the worldvolume of D-branes in the type IIA
superstring theory, where additional D-branes and NS5-branes are
used in order to get the required amount of supersymmetry, field
content and superpotential. This intersecting branes picture
provides at low energy and small string coupling limit the
classical gauge field theory. One way of analyzing the quantum
properties of the system is by lifting a type IIA brane
configuration to M theory and realizing it using an M5-brane
wrapping a curve. This method has been very successful for
analyzing the quantum vacua structure of supersymmetric gauge
theories. In these cases the M5-brane is wrapping a holomorphic
curve, whose properties encode the supersymmetric vacua structure.
This works despite the fact that the M5-brane description is valid
for large string coupling, which is the opposite limit to that of
the gauge theory one. The reason for this success is the
holomorphicity property of the quantities being studied. Indeed,
non-holomorphic quantities, such as the Kahler potential and
higher derivative couplings, differ between the gauge theory and
the M5-brane description \deBoerZY.

When supersymmetry is broken, the M5-brane is wrapping a
non-holomorphic curve of minimal volume (see e.g.
\WittenEP\deBoerBY). However, in this case there is no reason for
an agreement between the gauge field theory and the M5-brane
description, since their regimes of validity are very different.
Such a disagreement was found, for instance, between the quantum
ISS model and the M5-brane description \nostring\nostringa. Other
examples have been studied in
\clastring\clastringa\AhnYQ\twobe\twobea\twobeb. \foot{In this
M5-brane framework one can study also the brane/antibrane
configurations in type IIA, or their type IIB dual, which usually
do not have a gauge theory limit
\nogauge\nogaugea\nogaugeb\nogaugec\MarsanoFE\beem.}

It is clear that if one is interested in the quantum properties of
the gauge field theory, the way to proceed is to analyze the
intersecting branes configuration in the gauge theory limit. A
different study is to analyze supersymmetry breaking in the
framework of an M5-brane wrapping a curve. This theory is a
six-dimensional one at high-energy and a four-dimensional one at
length scales much larger than the typical size of the curve. In
this paper, we will analyze supersymmetry breaking in this
framework.

As noted above, supersymmetric vacua are realized as an M5-brane
wrapping a holomorphic curve. One may define spontaneous
supersymmetry breaking vacua as an M5-brane wrapping a
non-holomorphic minimal volume curve, which has holomorphic
boundary conditions at infinity. Thus, the curve has the same
asymptotics as that of a supersymmetric one, but differs in the
interior. One may also define an explicit breaking by an M5-brane
wrapping a non-holomorphic minimal volume curve, which has
non-holomorphic boundary conditions at infinity. Note that these
definitions are motivated by the four-dimensional gauge field
theory. From the M5-brane theory viewpoint, different minimal
volume curves are different choices of vacua, while the
high-energy six-dimensional worldvolume theory is supersymmetric.

In this paper we will consider intersecting brane configurations
realizing ${\cal N}=2$ supersymmetric gauge theories broken to
${\cal N}=1$ by multitrace superpotentials, and softly to ${\cal
N}=0$. We will analyze in the M5-brane framework the analogs of
spontaneous supersymmetry breaking and soft supersymmetry
breaking.

\subsec{Summary of the results}

We will start by presenting in Section 2, following the work of
\OoguriIU\PastrasQR, the field theory analysis of pure ${\cal
N}=2$ SYM with gauge group $G$, broken to ${\cal N}=1$ by the
higher trace superpotential for the adjoint
 \eqn\multitr{
 W=\sum_{i=1}^ks_i\Tr\, \Phi^{i+1} \ ,
 }
for $k>{\rm rank}\,G$. For a particular choice of couplings $s_k$,
the gauge theory develops a long-lived meta-stable vacuum at the
origin of the ${\cal N}=2$ Coulomb branch. The existence of this
vacuum relies on the exact knowledge of the ${\cal N}=2$ Kahler
potential.\foot{An example of a metastable vacuum in ${\cal N}=2$
gauge theory with flavors and a FI term has been studied in
\AraiMD.}

In Section 3 we will construct the type IIA brane configuration
that realizes the classical gauge theory \multitr\ by taking $k$
NS5 branes at an angle and suspending D4-branes between them, with
more NS5 branes than D4-branes. We will explicitly work out the
difference between an $SU(N)$ and $U(N)$ gauge theory. This
difference is important since the $N$ D-branes worldvolume gauge
group is $U(N)$ and, when $k>{\rm rank}\,G$, the abelian factor,
corresponding to the center of mass of the D4-brane stack, plays a
fundamental role. The superpotential \multitr\ gives rise to $k$
extra supersymmetric brane configurations. We will lift to M
theory these supersymmetric vacua, by considering an M5-brane
wrapping a holomorphic curve. These are new vacua whose lift is
not part of the analysis in \HoriAB\deBoerAP.
 Because $k>{\rm rank}\,G$,
the M5-brane  has several disconnected components, and it
successfully reproduces all the gauge theory supersymmetric vacua.

In Section 4, we will consider the meta-stable supersymmetry
breaking vacuum in the M5-brane framework. We will look for
non-holomorphic minimal volume curve, with holomorphic asymptotic
boundary conditions. We will see that there is no such curve, even
when we take into account the gravitational backreaction of the
disconnected branches of the M5-brane. Thus, the quantum
meta-stable gauge theory vacuum is not reproduced in the M5-brane
picture.

In Section 5, we will allow the M5-brane curve to have
non-holomorphic boundary conditions at infinity of the kind
 \eqn\boundarywv{
 w=m(v+\bar v) \ .
 }
We will find a family of minimal area non-holomorphic genus one
curves, whose boundary conditions are parameterized by the modulus
$\tau$ of the torus, each of which provides a lift of the
intersecting branes configurations. We evaluate the action of the
M5-brane wrapping this curve, which represents the energy of these
vacua. In Section 6 we interpret the boundary condition
\boundarywv\ as the analog of soft supersymmetry breaking in the
${\cal N}=2$ gauge theory perturbed by
 \eqn\softbre{
 {\cal L}_{soft}=\int d^4\theta {X^\dagger X\over \Lambda_s^2}u_1^\dagger
 u_1+\int d^2\theta M\, u_1^2+{\rm h.c.}
 }
where $u_1=\Tr\, \Phi$ is part of the visible sector, $X$ is the
hidden sector and $M$ is a spurion superfield.

There are three appendices in which we collect some useful
formulae on elliptic functions, we give a parametric description
of the ${\cal N}=2$ curve and we provide the details of the
solution to the non-holomorphic minimal area equations.

\newsec{Gauge theory analysis: multitrace deformations}

 We will review ${\cal N}=2$ gauge theory
broken to ${\cal N}=1$ by a superpotential for the adjoint chiral
superfield and describe its supersymmetric vacua and its scalar
potential. A particular choice of superpotential leads to the
existence of local minima of the scalar potential, which are
metastable vacua that dynamically break supersymmetry. This has
been discussed for $SU(2)$ gauge group in \PastrasQR\ and for
generic $SU(N)$ gauge group in \OoguriIU.

We consider ${\cal N}=2$ supersymmetric gauge theories with $U(N)$
gauge group. We will need $U(N)$ rather than $SU(N)$ gauge group
because the former is naturally realized by the brane
configurations. The chiral ring of the $U(N)$ gauge theory is
generated by $u_r={1\over r}\langle \Tr \Phi^{r}\rangle$, for
$r=1,\ldots,N$, where $\Phi$ is the adjoint chiral superfield in
the ${\cal N}=2$ gauge supermultiplet. If we denote by $a_i$ the
classical eigenvalues of the adjoint, then classically we have
 \eqn\chiraltwo{
 u_r=\sum_{i=1}^N a_i^r \ ,
 }
and the $u_r$ parameterize the moduli space of the Coulomb branch
of the ${\cal N}=2$ gauge theory, for $r=1,\ldots,N$. At a generic
point on the moduli space, the gauge symmetry is broken to its
maximal abelian subgroup $U(1)^{N}$ and the theory is in the
Coulomb branch. The chiral ring is conveniently encoded in the
characteristic polynomial $P_{N}(v,u_r)= \det (v-\Phi)=v^{N}
\exp{\left(-\sum_{r=1}^\infty {u_r\over
 v^r}\right)}$. Since $P_{N}(v)$ is a degree ${N}$ polynomial in $v$, we need to
impose that the coefficients of the negative powers in the Laurent
expansion vanish. In this way we can express the higher trace
operators $u_{r>{N}}$ in terms of the first ${N}$ operators
$u_{r\leq {N}}$. The ${\cal N}=2$ gauge theory physics is
described at low energy by the hyperelliptic curve
$y^2=P_{N}(v)^2-4\Lambda^{4{N}}$. The expectation values of the
chiral ring operators can be read from the curve as
 \eqn\vevchi{
 u_r=\oint_{\infty} dv {v^rP_{N}'(v)\over y} \ .
 }
The first ${N}$ operators in \chiraltwo\ are exact: they do not
receive quantum corrections in the ${\cal N}=2$ theory, but when
$r>{N}$ they get quantum corrections.

Now we break ${\cal N}=2$ supersymmetry to ${\cal N}=1$ by adding
a tree level superpotential
 \eqn\superpot{
 W=\sum_{r=0}^{k}s_ru_{r+1} \ ,
 }
where we will be interested in particular in the case where $k>
{N}$. The higher trace operators $u_r$ are to be understood as
multitrace interactions, when written in the usual basis of the
first $N$ chiral ring operators.\foot{At low energy in the ${\cal
N}=2$ theory, the gauge dynamics of the $U(1)$ part is frozen, so
the ${\cal N}=2$ theory (without the superpotential \superpot) is
effectively $SU({N})$. However, when we add the interaction
\superpot, the $U(1)$ part of the adjoint chiral superfield
interacts with the remaining $SU({N})$ part through the Yukawa
couplings, hence we cannot disregard the $U(1)$ part of the
dynamics, that will be crucial to identify the correct vacua in
the brane picture.} Let us briefly discuss the vacuum structure of
these gauge theories. The ${\cal N}=2$ theory has a quantum moduli
space of supersymmetric vacua parameterized by the $u_r$ for
$r=1,\ldots,N$. When we add the superpotential \superpot, the
moduli space is lifted to a discrete set of supersymmetric vacua,
given by the $u_r$ in \chiraltwo\ where the eigenvalues of the
adjoint are at the roots of
 \eqn\eoms{
 W'(v)=s_k\prod_{i=1}^k(v-a_i) \ ,
 }
modded out by the Weyl reflection, so the number of classical
vacua in the gauge theory is
 \eqn\numberva{ \pmatrix{N+k-1\cr N} \ .
}

The non-supersymmetric vacua are the non-zero minima of the scalar
potential
 \eqn\scalarpot{
 V=g^{i\bar j}\partial_iW\,\partial_{\bar j}W \ ,}
where $g_{i\bar j}$ is the Kahler potential of the ${\cal N}=1$
gauge theory. In general it is difficult to compute the ${\cal
N}=1$ Kahler potential, however, in the regime where the
superpotential is just a small perturbation, we can reliably use
the ${\cal N}=2$ Kahler metric on the moduli space
 \eqn\kahlerme{
 g_{r\bar s}={\rm Im}\, \tau_{i j}{da^i\over du_r}{{d\bar a^j\over
 d\bar u_s}} \ ,
 }
where $\tau_{ij}$ is the matrix of the low energy $U(1)$
couplings. The authors of \OoguriIU\ showed that any point on the
${\cal N}=2$ moduli space of vacua can be lifted to a
non-supersymmetric metastable vacuum by an appropriate choice of
superpotential \superpot\ with higher trace operators. In
particular, if we integrate out the $u_1$ modulus so that we are
left with an $SU(N)$ gauge group, one can lift the origin of the
$SU(N)$ moduli space by turning on the tree level superpotential
 \eqn\origin{
 W=\lambda\left({u_N\over N}+{(N-1)^2\over6N^3}{u_{3N}\over
 \Lambda^{2N}}\right) \ ,
 }
where $u_N, u_{3N}$ are the $SU(N)$ operators and $\lambda$ is a
small coupling. Moreover, the metastable vacuum at the origin can
be made parametrically long lived against decays to both the
classical supersymmetric vacua and the quantum vacua at the points
where dyons condense, by appropriately tuning the couplings and
the dynamical scale.

\subsec{Metastable vacua with $U(2)$ gauge group}

Let us work out in more detail the case of $U(2)$ gauge group,
that will be relevant for the brane configuration. The ${\cal
N}=2$ chiral ring is generated by $u_1$ and $u_2$. If we split the
classical $U(2)$ adjoint chiral superfield into its $U(1)$ part
and its $SU(2)$ part as $\Phi=\Id x+\varphi$, we can express the
modulus of the $SU(2)$ gauge group $u=\half\Tr\varphi^2$ as $
u=u_2-{1\over4}u_1^2$. Therefore, the origin $u=0$ of the $SU(2)$
moduli space occurs at $u_2=u_1^2/4$ in the $U(2)$ theory.

Let us first add a tree level mass term for the adjoint, namely
$W=mu_2$. In the regime of small mass $m$ we can compute the exact
scalar potential, which is simply $V=g^{u_2\bar u_2}\vert
m\vert^2$ with the metric $g_{u_2\bar u_2}$ given in \kahlerme. As
explained above, the overall $U(1)$ part does not contribute to
the ${\cal N}=2$ dynamics, that determines the Kahler potential:
the metric for the $u_2$ modulus is thus same as the metric for
the modulus $u$ of the $SU(2)$ gauge theory. Hence, as far as the
computation of the scalar potential is concerned, we can integrate
out $u_1$ upon its equations of motion and compute $V$ using the
effective superpotential for $u_2$. The scalar potential in the
massive case is depicted in Fig. 1a. It has an extremum at the
origin $u_2=0$, however it is a saddle point. In Section 4 we will
argue why naively one may expect to see this extremum in the brane
picture, since it might correspond to a solution to the M theory
equations of motion. However, the actual M theory computation will
show that there is no such solution at all.

We would like to study the metastable supersymmetry breaking
vacuum found in \OoguriIU\PastrasQR. Let us introduce the
superpotential
 \eqn\superoo{
 W=s_1u_2+s_5u_6\ ,
 }
whose equations of motion can be written as
 \eqn\eomsoo{\eqalign{
 u_1u_2(2u_2-u_1^2)=&\,0 \ ,\cr
 s_1+s_5\left(u_2^2+2u_1^2u_2-{1\over4}u_1^4\right)=&\,0 \ ,
 }}
where we expressed $u_6$ in terms of the $u_1$ and $u_2$.

We have six solutions for $u_1$ to be integrated out, giving an
effective potential for $u_2$
 \eqn\effectW{\eqalign{
 u_1=0\qquad &\Rightarrow W^{(1)}_{eff}=s_1u_2+{s_5\over3} u_2^3 \ ,\cr
 u_1=2u_2 \qquad &\Rightarrow W^{(2)}_{eff}=s_1u_2+{4s_5\over3} u_2^3 \ ,\cr
 u_1^2=\pm\sqrt{4s_1/s_5} \qquad &\Rightarrow W^{(3)}_{eff}=\pm -2 \sqrt{s_1s_5}
 u_2^2+{s_5\over3} u_2^3 \ .
 }}
The scalar potential $V^{(i)}(u_2)=g^{u_2\bar u_2}\vert
\partial_{u_2}W^{(i)}_{eff}\vert^2$ will have three different expressions on the three
different branches in \effectW. The analysis in each branch
reduces then to the one in \OoguriIU\PastrasQR\ and it turns out
that $V^{(1)}$ and $V^{(2)}$ display a metastable vacuum at the
origin of the $u_2$ moduli space in a special range of the
coupling $s_5/s_1$, $\lambda^{(i)}_-<{s_1\over
s_5}<\lambda^{(i)}_+$ where
$\lambda^{(1)}_\pm=1/24\pm\left({\Gamma(3/4)\over2\Gamma(5/4)}\right)^4$
and $\lambda^{(2)}_\pm=4\lambda^{(1)}_\pm$. The metastable vacuum
is shown in Fig. 1b. \ifig\figone{Plot of the scalar potential. In
Fig.1a, the superpotential $W=mu_2$ gives a saddle point at the
origin. In Fig.1b, the superpotential $W^{(1)}_{eff}$ in \effectW\
transforms the saddle point into a local minimum.}
{\epsfxsize2.6in\epsfbox{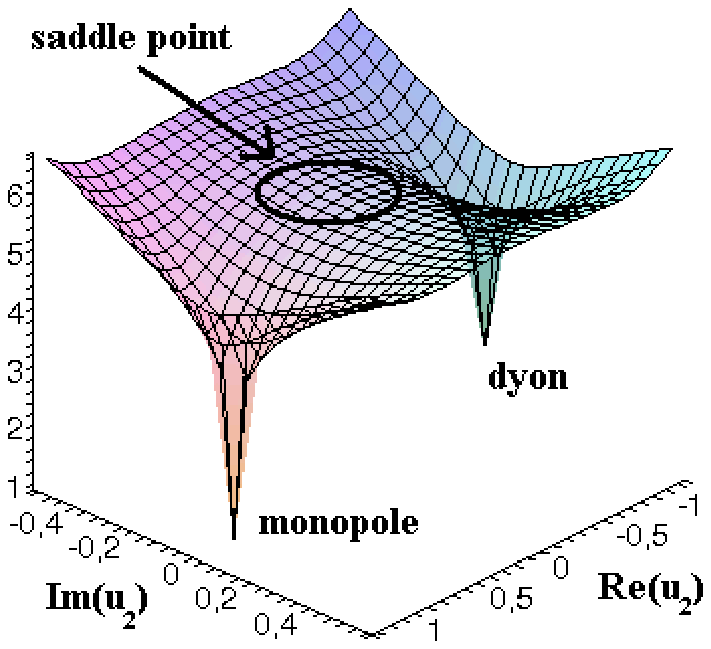}\epsfxsize2.7in\epsfbox{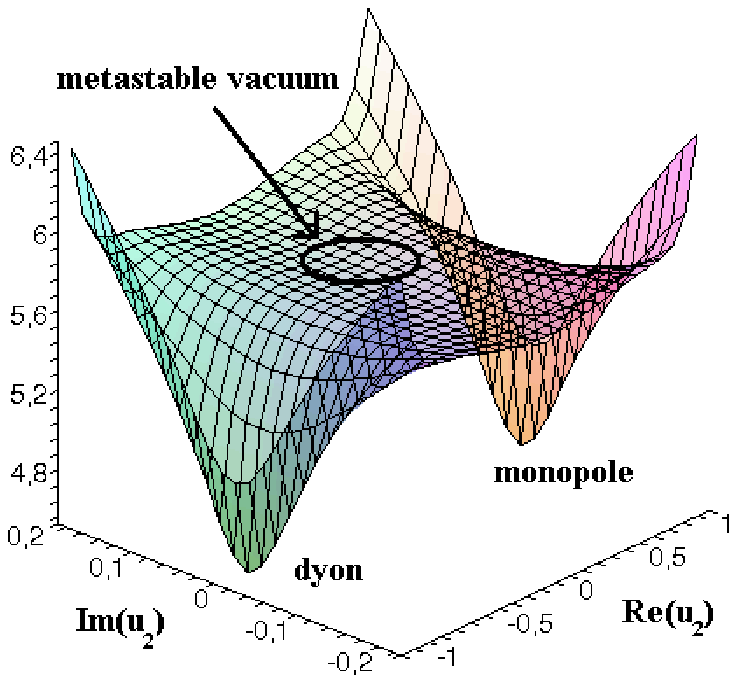}}

Let us comment on the physics of the classical supersymmetric
vacua \eomsoo. They correspond to generic values of the moduli
$u_1$ and $u_2$. Close to the origin of the moduli space, once we
integrate out the dynamics of the abelian factor corresponding to
$u_1$, the softly broken $U(2)$ gauge theory admits an effective
description in terms of an abelian gauge theory coupled to two
chiral superfields $M$ and $\widetilde M$, representing magnetic
monopoles, whose superpotential is
 \eqn\magnetic{
 \widetilde{W}=\widetilde M A M+\sum_{i=1}^k s_i u_i \ ,
 }
where the last term is \effectW. At a generic point on the moduli
space, the equations of motion of \magnetic\ set $\widetilde M=
M=0$: the monopoles are massive, so the curve is not degenerate.
In addition, there are two extra supersymmetric vacua where a
monopole or a dyon condenses and the curve degenerates.

The reason for the existence of these metastable vacua is the
following \OoguriIU. The metric \kahlerme\ on the moduli space of
the ${\cal N}=2$ Coulomb branch has positive definite curvature
almost everywhere. There exists therefore a suitable
superpotential such that any point on the ${\cal N}=2$ moduli
space can be lifted to a metastable vacuum. Around any regular
point on the moduli space, one can go to the coordinate system
$z^i$, for $i=1,\ldots,N$, adapted to that point. Then, it is
generically possible to choose a superpotential cubic in $z^i$,
such that the scalar potential \scalarpot\ has a local minimum at
the origin, in this coordinate system. Higher powers than cubic in
general do not affect the metastability of the vacuum, and in fact
one can add such irrelevant terms as long as their couplings are
small. The analysis is valid when the superpotential is treated as
a small perturbation, so that one can trust the ${\cal N}=2$
Kahler metric.

\newsec{Supersymmetric vacua in the brane picture: multitrace deformations}

We have discussed how the gauge theory with the multitrace
superpotential \origin\ develops a metastable vacuum at the origin
of the ${\cal N}=2$ moduli space. On top of that, the multitrace
deformation \origin\ gives rise to a large number of
supersymmetric vacua. In this Section we will discuss the type IIA
description of these gauge theory supersymmetric vacua and their
lift to M theory. Due to the fact that the degree of the
superpotential is larger than the number of color, the M theory
lift will be different from the ones studied in the past (for a
review see \GiveonSR\ and references therein), where the degree of
the superpotential was taken to be at most equal to the rank of
the gauge group. The new ingredient is that the ``excess" $k-N$ NS
fivebranes, once lifted to M theory, become a bunch of
disconnected components of the M5-brane worldvolume.

 \subsec{Type IIA setup}

Let us consider type IIA string theory in flat ten dimensions. The
brane configuration describing the classical ${\cal N}=1$ gauge
theory with degree $k+1$ superpotential \superpot\ consists of one
fivebrane NS, $k$ fivebranes NS' and $N$ D4-branes, whose
worldvolumes extend along
 \eqn\braneconfig{
\matrix{    & x_0     & x_1    & x_2 & x_3 & v & x_6 & x_7 & w\cr
          NS & \bullet & \bullet &\bullet &\bullet  &\bullet
          &\times &\times &\times \cr
          NS\,'& \bullet & \bullet &\bullet &\bullet  & /
          &\times &\times &/ \cr
          D4& \bullet & \bullet &\bullet &\bullet  & \times
          &\bullet &\times &\times \cr
           }
 }
where $v=x_4+ix_5$ and $w=x_8+ix_9$. The $k$ NS' branes are
rotated in the $(v,w)$ directions and stuck at a point in $x_6$.
The gauge theory eigenvalues of the $U(N)$ adjoint $\Phi$
correspond in the brane picture to the positions of the D4-branes
along the $v=x_4+ix_5$ direction at $w=0$. In particular, the
$U(1)$ part $u_1/N=\Tr \Phi/N$ of the adjoint represents the
center of mass coordinate of the system of the D4-branes, while
the operators $u_r$ for $r=2,\ldots,N$ parameterize the relative
displacement of the D4-branes in the $v$-plane. When the $k$ NS'
are rotated in the $(v,w)$ direction, their position along the $v$
direction at $w=0$ is given by the solutions of the classical
equations of motion in which all the D4-branes are on top of each
other, namely for $u_{r>1}=0$ and $u_1\neq0$. In particular, given
a generic superpotential $W(\Phi)$ in \eoms, the $k$ NS' branes
intersect the plane $w=0$ at $v=a_i$. The number of ways to
suspend the $N<k$ D4-branes between the NS and the $k$ NS' is
precisely \numberva, showing the one to one correspondence with
the classical supersymmetric vacua of the gauge theory.

As an illustrative example, consider the classical $U(2)$ gauge
theory superpotential \superoo\ responsible for the metastable
vacua in Fig. 1b. The positions of the $k=5$ NS' branes are
determined by $W_{eff}^{(1)}$ in \effectW, and we have drawn their
locations in the $v$ plane in Fig. 3. Note that in order to
reproduce correctly the vacua it is crucial to take into account
the Yukawa couplings between the $U(1)$ modulus $u_1$ and the
nonabelian part of the adjoint superfield. \ifig\utwons{The
positions of the $k=5$ NS' branes in the $v=x_4+ix_5$ plane at
$w=0$, corresponding to the gauge theory vacuum $W_{eff}^{(1)}$
\effectW.} {\epsfxsize1.4in\epsfbox{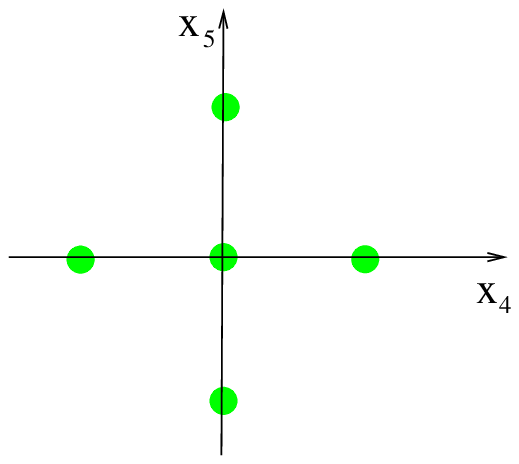}}

\subsec{M theory lift and disconnected curves}

We want to discuss the lift to M theory of the classical gauge
theory vacua, which are in one to one correspondence to the
classical brane configurations. The ${\cal N}=2$ supersymmetric
theory has a moduli space of vacua. It corresponds to parallel NS
and NS' branes, extended along the $v$-plane at $w=0$. In this
case the D4-branes are free to move in the $v$-direction, and
their positions parameterize the Coulomb branch of the gauge
theory. When we add an ${\cal N}=1$ superpotential to the gauge
theory, the moduli space is lifted, leaving just an isolated
number of vacua. We have $N$ isolated vacua, corresponding to the
points at which a massless monopole condenses, that in the low
energy theory represent the $N$ gaugino condensate vacua. In
addition to that, we have more supersymmetric vacua, given by the
solution to the F-term equations.

Consider the vacuum in which each of the $N$ D4-branes is attached
to a different NS' brane. The $k$ NS' branes are located at the
roots of \eoms. We separate them in two sets: to the first $N$ of
them, that intersect $w=0$ at the positions $v=a_1,\ldots,a_N$, we
attach the $N$ D4-branes; the remaining $k-N$ NS' are just
spectators, and we place them at the positions
$v=a_{N+1},\ldots,a_k$. When we switch on the type IIA string
coupling $g_s<<1$, the eleventh dimensional circle $x_{10}$ opens
up. As usual we introduce a new complex coordinate as
$t=\exp\left[-(x_6+ix_{10})/R\right]$, where $R$ is the M theory
radius. Quantum mechanically, a D4-brane ending on the NS5 brane
bends it at infinity. The classical brane configuration we have
just described consists then of three different asymptotic regions
in M theory as shown in Fig. 3. The first region is at
$x_6=-\infty$, that is $t=\infty$, where the NS brane is bent by
all the $N$ D4-branes attached to it
 \eqn\firstasy{
 t\sim\infty,\quad v\sim\infty:\qquad\matrix{t\sim 2 v^N\cr
 w\sim s_k\Lambda^{2N}/ v^N}.
 }
The second asymptotic region is at $x_6=\infty$, that is $t=0$,
where we have $N$ NS' branes, which are bent by the $N$ D4-branes
attached to them. Each NS' is rotated in the $(v,w)$ plane, so
that at infinity we need $N$ different solutions for $w$ as a
function of $v$
 \eqn\secondasy{
 t\sim0,\quad v\sim\infty:\qquad\matrix{t\sim 2\Lambda^{2N}/ v^N\cr
 w\sim s_k\prod_{i=1}^N(v-a_i)}.
 }
The third asymptotic region corresponds to the $k-N$ spectator NS'
branes, to which no D4-brane is attached. Since they feel no
force, their worldvolume is flat and extend at an angle in the
$(v,w)$ direction and at a fixed position $t=t_0$
 \eqn\thirdasy{
 t\sim t_0,\quad v\sim\infty:\qquad
 w\sim s_k\prod_{i=N+1}^k(v-a_i).
 }
\ifig\lift{The small $g_s$ description of the $U(2)$ gauge theory
with $k=5$. The NS' branes are at an angle $\theta$ in the $(v,w)$
plane. The brane configuration corresponds to a vacuum with a
nonzero vev for both $u_1$ and $u_2$ moduli. The asymptotic region
\firstasy\ corresponds to the red NS brane on the right; the
asymptotic region \secondasy\ corresponds to the brown NS' branes
on the right; the third asymptotic region \thirdasy\ corresponds
to the flat blue NS' branes. The green lines are the two D4-branes
suspended between the NS and the NS' fivebranes. }
{\epsfxsize3.9in\epsfbox{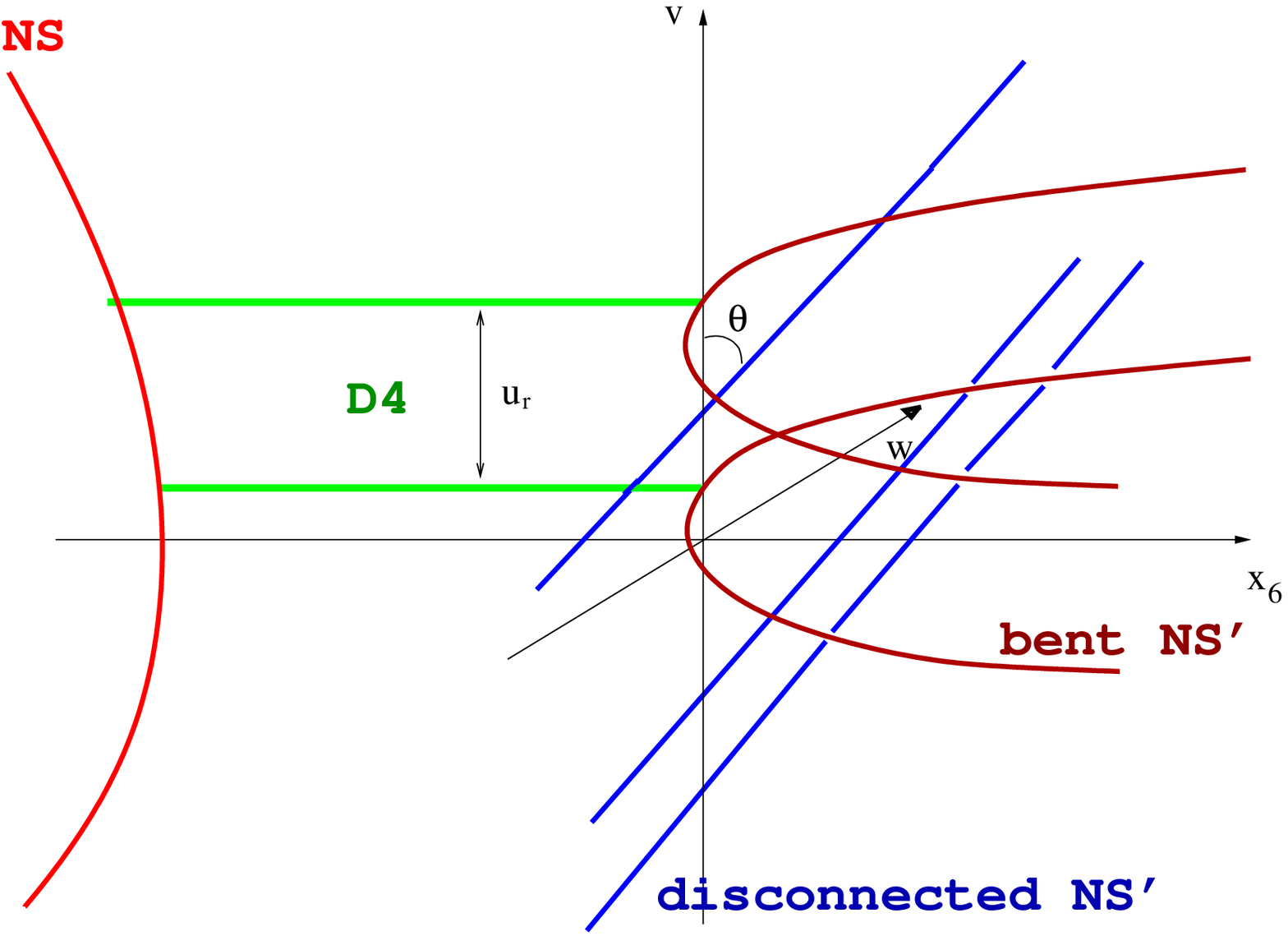}}

The M theory configuration that satisfies these three asymptotic
boundary conditions is a fivebrane with worldvolume $R^{1,3}\times
\Sigma$, where
 \eqn\curve{
 \Sigma=\Sigma_c\cup\Sigma_d \ ,
 }
is a holomorphic curve consisting of two disconnected components.
In the case in which each D4 brane ends on a different NS' brane,
the component $\Sigma_c$ of the fivebrane satisfying the first and
second boundary conditions \firstasy-\secondasy\ is given by
 \eqn\concurve{
 \Sigma_c:\qquad \left\{\matrix{&v=
 \left({t\over2}\right)^{1\over N}+\Lambda^2\left({2\over t}\right)^{1\over N}\ ,\cr
 &w=s_k\prod_{i=1}^N(v-a_i)\ . }\right.
 }
The first equation is the usual Seiberg-Witten curve for the
$U(N)$ gauge theory at the point in the moduli space in which it
degenerates to a sphere. The second disconnected component
$\Sigma_d$ of the fivebrane worldvolume simply consists of the
collection of the spectator flat $k-N$ NS' branes and is given
by\foot{The general case, in which multiple D4 branes end on each
NS' branes, is given by a partial degeneration of the ${\cal N}=2$
curve. It is discussed in Eq. (4.21) of \deBoerAP. The
disconnected part $\Sigma_c$ can be easily obtained as well, as a
collection of the leftover disconnected NS' branes.}
  \eqn\discurve{
 \Sigma_d:\qquad \left\{\matrix{&t=t_0\ ,\cr
 &w=s_k\prod_{i=N+1}^k(v-a_i)\ ,}\right.
 }

\newsec{M5-brane non-supersymmetric vacua: no metastable spontaneous breaking}

In this Section we will discuss the issue of metastable
spontaneous supersymmetry breaking in the framework of an M5-brane
wrapping a non-holomorphic curve.

Recall that in gauge theory a metastable supersymmetry breaking
vacuum is realized as follows: one introduces a  supersymmetric
lagrangian and computes  the scalar potential \scalarpot. A local
minimum with non-zero energy breaks  supersymmetry spontaneously.
If there are other minima at lower energies, then the
supersymmetry breaking minimum is metastable towards tunnelling to
these other vacua, and  in order to be phenomenologically
interesting, it must be long-lived, i.e. its decay to the other
lower energy minima being parametrically small.

We may translate this discussion to the M5-brane framework, by
defining a spontaneous breaking of supersymmetry as a wrapping of
a non-holomorphic minimal volume curve with holomorphic boundary
conditions at infinity. These holomorphic boundary conditions
correspond to the holomorphic classical superpotential in the
gauge theory. A stable non-supersymmetric minimum such as the IYIT
model \IntriligatorPU, realized on the branes in \deBoerBY, will
translate to having no holomorphic curve and only a
non-holomorphic minimal volume curve with holomorphic boundary
conditions. On the other hand, having both a supersymmetric vacuum
and a non-supersymmetric one corresponds to two different
solutions to the minimal volume equations with same holomorphic
boundary conditions: one holomorphic curve, corresponding to the
supersymmetric vacuum, and one non-holomorphic curve,
corresponding to the metastable non-supersymmetric vacuum.

In the case of the metastable vacuum found in ${\cal N}=1$ SQCD
with massive flavors \ISS, it has been shown that the M5-brane
theory does not realize the gauge theory metastable vacuum
\nostring. As we noted before, this is not unexpected, since the
M5-brane framework regime of validity and the gauge theory one are
not the same. Indeed, here as well we will see that the metastable
vacuum in softly broken ${\cal N}=2$ gauge theory
\OoguriIU\PastrasQR\ are not realized on the worldvolume of the
M5-brane. Once we fix holomorphic boundary conditions at infinity,
we find only holomorphic minimal volume curves.

\subsec{The minimal volume equations}

The worldvolume of the M5-brane is $R^{1,3} \times \Sigma$, where
$\Sigma$ is a two-dimensional curve. If we consider a Nambu-Goto
form for the bosonic part of the action, then the area of the two-
dimensional curve plays the role of the potential energy
\foot{Since the curves are non-compact this area is infinite, it
needs to be regularized, as we discuss later on.}
 \eqn\areac{
 {\rm Area}(\Sigma_g)=\int_{\Sigma_g}d^2x\,\sqrt{g} \ ,
 }
where $g$ is the determinant of the induced metric on the
worldvolume. The area element can be expressed as
 \eqn\areael{
 \sqrt{g}d^2x=g_{z\bar z}d^2z=G_{i\bar j}\left(
 \partial_zX^i\partial_{\bar z}X^{\bar j}+\partial_{\bar
 z}X^i\partial_{z}X^{\bar j}\right)d^2 z \ ,
 }
where $G_{i\bar j}$ is the spacetime metric. The equations of
motion are then equivalent to solving for a minimal area surface,
namely the embedding coordinates $X^i$ and $X^{\bar j}$ must
satisfy
  \eqn\eomsem{
 G_{i\bar j}\partial_{\bar z}\partial_z X^i+\partial_{\bar
 z}X^i\partial_z X^k{\partial\over \partial X^k}G_{i\bar j}=0 \ ,
 }
and the Virasoro constraint
 \eqn\viraco{
 G_{i\bar j}\partial_zX^i\partial_z X^{\bar j}=0 \ .
 }
In our setup, the spacetime embedding coordinates are
$X^i=(w,v,s),X^{\bar{i}}=(\bar{w},\bar{v},\bar{s})$. When the
spacetime metric is flat, the second term in \eomsem\ drops.

We would like to find embedding coordinates $(w,v,s)$, which
satisfy the M5-brane equations of motion and Virasoro constraints.
The first condition is
 \eqn\eomfive{
 \partial\bar\partial s=\partial\bar\partial
 v=\partial\bar\partial w=0 \ ,
 }
which is solved by harmonic functions
 \eqn\harm{\eqalign{s(z,\bar
z)=&s_H(z)+\overline{s_A(z)} \ , \cr v(z,\bar
z)=&v_H(z)+\overline{v_A(z)} \ , \cr w(z,\bar
z)=&w_H(z)+\overline{w_A(z)} \ . \cr
 }}
The Virasoro constraint \viraco\ reads
 \eqn\vira{
 g_s^2\partial s_H\partial s_A+\partial v_H\partial v_A+\partial
 w_H\partial w_A=0 \ ,
 }
where the $g_s$ factor comes from the metric. Note that a
holomorphic curve automatically satisfies both equations of motion
\eomfive\ and \harm.

\subsec{Breaking ${\cal N}=2$ to ${\cal N}=1$ in the brane
picture}

Let us first recall the parametric description of the ${\cal N}=2$
holomorphic curve for $U(2)$ gauge theory in terms of a torus with
coordinate $z$ and period $\tau$ as in Fig.4. More details are
given in the Appendix B. The embedding coordinates at a point
where $U(2)$ is broken to $U(1)\times U(1)$ are given by
 \eqn\embetwo{\eqalign{
 s_{SW}(z)=&2(F(z-a_1)-F(z-a_2)-\pi i z) \ ,\cr
 v_{SW}(z)=&A(F^{(1)}(z-a_1)-F^{(1)}(z-a_1)-i\pi)+\half u_1 \ ,\cr
 w_{SW}(z)=&0 \ ,
 }}
where the relation between $A,\tau$ and the usual moduli and
dynamical scale are derived in Appendix B. We have introduced the
function $F(z)=\ln\theta_3[\pi(z-\tilde\tau)]$, where $\tilde
\tau=(\tau+1)/2$, and denoted its derivative by $F^{(1)}(z)$. The
properties of this function are discussed in Appendix A, following
the conventions in \MarsanoFE. The embedding coordinates satisfy
the following boundary conditions at the NS and NS' branes
 \eqn\masso{
 NS:\,\,z\sim a_2\qquad \left\{\matrix{&w\sim 0\ ,\cr
 &v\sim\infty \ ,\cr
 & t\sim 2v^N \ .    }\right. \qquad
 NS\,':\,\,z\sim a_1\qquad \left\{\matrix{&w\sim 0 \ ,\cr
 &v\sim\infty \ ,\cr
 & t\sim 2\Lambda^{2N}/v^N \ ,    }\right. }
where $t=e^{-s}$. In the formula \masso\ and in the following we
understand that $N=2$ and $N_1=N_2=1$, but we sometimes keep the
number of colors explicit.

When studying ${\cal N}=1$ holomorphic curves, we parameterized
the boundary conditions for the embedding coordinates $t=e^{-s}$
as in \firstasy\ and \secondasy, which in the ${\cal N}=2$ case
reduce to \masso. It is convenient to recast these asymptotics in
another form, which will be more appropriate when studying
non-holomorphic curves \MarsanoFE. These definitions are valid for
${\cal N}=2,1,0$. We specify the periods and residues of the
differential $ds$ on the torus: the residue at the location of the
NS brane is the total number of $D4$-branes, equal to the rank $N$
of the gauge group
 \eqn\resin{
 Res_{a_1}ds=Res_{a_2}ds=2\pi i N \ .
 }
The A-periods in the ${\cal N}=1$ language correspond to the ranks
of the low energy gauge groups
 \eqn\aper{
 \oint_{A_i}ds=2\pi i N_i \ ,
 }
which in the type IIA picture represents the number of D4-branes
that are piled up in the same stack, which on the M theory side it
is the number of times the M5-brane wraps the eleventh dimensional
circle. In our case $N_i=1$, since we are on the ${\cal N}=2$
Coulomb branch (the $A_2$-period is defined up to the residue at
$a_1$). Then we have the constraint that the total $B$ period is
an integer. The compact $B$ periods are the differences in the
theta angle of consecutive low gauge groups, but in our case we
just have abelian gauge groups, so we fix it to zero
 \eqn\bconst{
 \oint_{B_1-B_2}ds=0.
 }
If we introduce a cutoff $v_0$ for $v$, when $z$ is close to the
marked points $a_1,a_2$, then the B-periods of $ds$ give the
four-dimensional running gauge couplings at the scale $v_0$
 \eqn\bper{
 \oint_{B_i}ds=2\pi i \alpha_i(v_0) \ ,
 }
where $\alpha(v_0)={\theta\over2\pi}+{4\pi i \over
g_{YM}^2(v_0)}$. In the ${\cal N}=2$ case \bper\  reproduces the
one loop part of the exact $\tau$
 \eqn\taunorun{
 \oint_{B_2}ds=-2\ln{v_0^2\over\Lambda^2}
 }
where $\Lambda$ is the ${\cal N}=2$ dynamical scale. In the
holomorphic case, fixing these periods of $ds$ gives back the
boundary conditions \masso.  The solution \embetwo\ satisfies the
various boundary conditions \resin, \aper. The constraint \bconst\
requires
 \eqn\bconre{ a\equiv a_2-a_1 =-{\tau\over2} \ .
 }
\ifig\gammatt{The parametric description of the torus in the $z$
plane. The marked points at $z=a_1,a_2$ are the location of the
NS' and NS fivebranes. Their distance is fixed to
$a_2-a_1=-\tau/2$. The cycles $B_1$ and $B_2$ are non-compact.}
{\epsfxsize3in\epsfbox{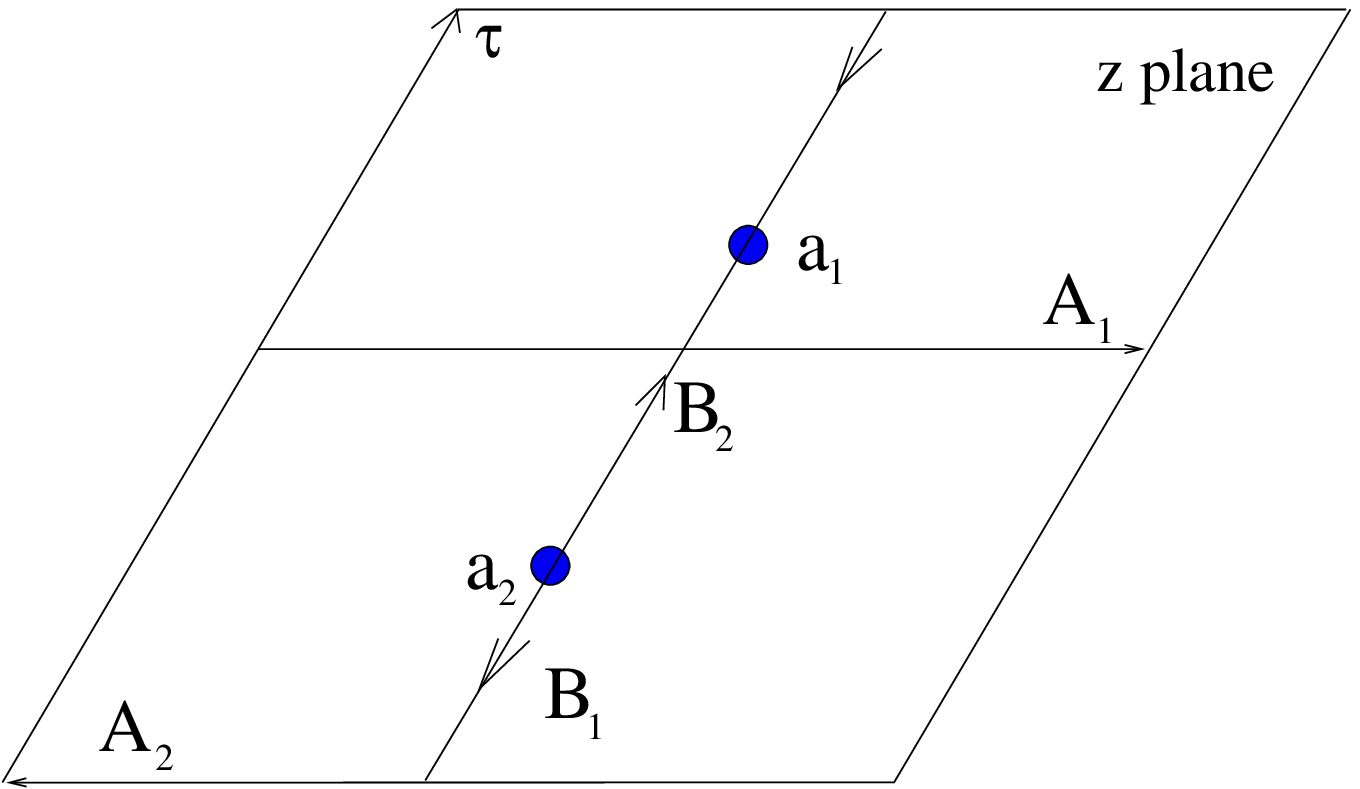}}

Let us break now the ${\cal N}=2$ supersymmetry to ${\cal N}=1$ by
the superpotential
 \eqn\massive{
 W=mu_2={m\over 2}\Tr \Phi^2 \ . }
The corresponding boundary conditions \massive\ are usually taken
to be \masso\ at the NS brane, namely $w=0$, while at the NS'
brane one takes $w=mv$. For later convenience, we perform a
rotation in the $(w,v)$ space and take more symmetric boundary
conditions given by
 \eqn\massns{
 NS:\,\,z\sim a_2\qquad \left\{\matrix{&w\sim -mv \ ,\cr
 &v\sim\infty \ ,\cr
 & t\sim 2\Lambda^4/v^2 \ .    }\right. \qquad
 NS\,':\,\,z\sim a_1\qquad \left\{\matrix{&w\sim mv \ ,\cr
 &v\sim\infty \ ,\cr
 & t\sim 2v^2 .    }\right.
 }
As expected from the gauge theory, there is no holomorphic torus
with holomorphic boundary conditions \massns. In fact, $w$ would
be an elliptic (i.e. doubly periodic) meromorphic function with
non-zero residue and this is not possible. Note that if, instead,
we look for a holomorphic curve with these boundary conditions but
with genus zero, then we find the holomorphic lift of the ${\cal
N}=2$ monopole and dyon points, where the torus degenerates to a
sphere \HoriAB. The latter are indeed supersymmetric vacua.

\subsec{M5-brane: no metastable spontaneous supersymmetry
breaking}

If we introduce the superpotential \massive, then the gauge theory
scalar potential $V(u_2)$, computed in the approximation of small
mass $m$, has an extremum (saddle point) at the origin $u_2=0$ as
we showed in Fig. 1a. In the following we ask whether there is in
the M5-brane framework a corresponding non-holomorphic curve with
asymptotically holomorphic boundary conditions \foot{In order to
distinguish a saddle point from a minimum one may study the
spectrum of fluctuations around the solution.}.

We will first look for a non-holomorphic minimal area torus with
boundary conditions \massns, corresponding to the massive gauge
theory \massive. Then, in the next Section we will consider the
multitrace deformation \superoo, that on the gauge theory side
gives rise to a metastable vacuum. As we have discussed in Section
3, this deformation is realized in the brane picture by adding
disconnected parts of the M5-brane worldvolume. We will attempt to
take the change in the brane configuration into account by
considering the effect of the gravitational interaction of the
disconnected curves on the part of the curve that in the type IIA
limit contains the D4-branes. Our analysis will show that the
gravitational interaction of the disconnected components can be
actually neglected and we are back to the first case. We will see
that the M5-brane does not exhibit the metastable
non-supersymmetric vacua.

Let us see what goes wrong if we try to lift to M theory the type
IIA intersecting brane configuration at the origin of the moduli
space. At $u_2=0$, the gauge symmetry in the quantum theory is
broken to $U(1)\times U(1)$, hence would-be the curve is a torus,
parameterized by the holomorphic coordinate $z$. The position of
each NS5 brane at infinity is a marked point on the torus at
$z=a_1,a_2$. This means that the surface is non-compact.

If we try to lift \massns\ with a non-holomorphic curve, we
encounter a problem. The equations of motion \harm\ imply that
$w,v$ and $s$ are harmonic and elliptic functions. We can achieve
this by adding to \embetwo\ and appropriate anti-holomorphic part
such that, in particular $w(z,\bar z)\sim \pm mv(z,\bar z)$ at
$z\to a_{1,2}$, and the functions are elliptic. However,  we have
to satisfy the Virasoro condition \vira\ and it is easy to see
that it is not possible to find such elliptic functions, not even
at first order in the small mass parameter $m$. This is because,
when we try to satisfy \vira\ in the vicinity of the marked points
$z\sim a_1,a_2$, the contribution from $w$ and $v$ contain a
fourth order pole, whose coefficient is always proportional to
$1+\vert m\vert^2$, that never vanishes. On the other hand, a
higher trace superpotential e.g. such as \superoo, corresponds to
boundary conditions of the form $w\sim v^k+\ldots$. This case is
even worse than the previous one. In fact, although it is possible
to have harmonic elliptic functions with these boundary
conditions, the leading contribution of $w$ to the Virasoro
condition \vira\ is now a pole of degree $2k+2$, while the leading
contribution of $v$ is still of fourth order and they do not
cancel.

\subsec{Backreaction of the disconnected components of the
fivebrane}

Let us introduce the higher trace deformation \superoo\ and
incorporate the gravitational interaction of the disconnected
components of the M5-brane. The eleven dimensional metric
$G_{i\bar j}$ is sourced by the $k-1$ parallel fivebranes, rotated
in the $(v,w)$ directions by an angle $\theta$, each of which
intersect the $v$-plane at $v=v_i$
 \eqn\metric{
 \eqalign{
 ds^2_{11}=&f^{-{1\over3}}dx_\parallel^2+f^{2\over3}(dr_\perp^2+r_\perp^2d\Omega_4^2)
 \ ,
  }}
 \eqn\harmo{
 f=1+\sum_{i=1}^{k-1}{c\over \vert r-r_i\vert^3},
 }
and the transverse coordinate reads
 \eqn\trans{
 \vert r-r_i\vert^2_\perp=x_7^2+\vert s\vert^2+\cos^2\theta \vert w\vert^2
 +\sin^2\theta(v-v_i)(\bar v-\bar v_i)-\sin\theta\cos\theta(\bar
 w(v-v_i)+w(\bar v-\bar v_i)) \ .
 }
The metric $G_{i\bar j}$ has the following non-zero components in
the $(v,w,s)$ directions
 \eqn\offmetric{\eqalign{
 G_{v\bar v}=& \cos^2\theta f^{-{1\over3}}+\sin^2\theta
 f^{2\over3}\simeq 1+\left(-{1\over3}+\sin^2\theta\right)\sum_i{c\over\vert r-r_i\vert_\perp^3} \ , \cr
G_{w\bar w}=& \sin^2\theta f^{-{1\over3}}+\cos^2\theta
 f^{2\over3}\simeq 1+\left(-{1\over3}+\cos^2\theta\right)\sum_i{c\over\vert r-r_i\vert_\perp^3} \ , \cr
G_{s\bar s}=&f^{{2\over3}}\simeq 1+{2\over3}\sum_i{c\over\vert
r-r_i\vert_\perp^3} \ , \cr
 G_{v\bar w}=& \sin\theta\cos\theta(
f^{-{1\over3}}-
 f^{2\over3})\simeq -\sin\theta\cos\theta\sum_i{c\over\vert r-r_i\vert_\perp^3}\ , \cr
 }}
where we expanded at large distance $r$ from the source NS'
branes. In order for the target space metric to be hermitian we
need to require $\theta=\bar\theta$.

Since we are looking at the solution to the equations of motion
and the Virasoro constraint \eomsem\ and \viraco\ in terms of
elliptic functions, we just need to evaluate these equations
around one pole, say $z\sim a_1$. The marked points on the torus
represent the location of the bent NS and NS' around infinity,
where the distance between the two bent fivebranes and the source
spectator fivebranes is very large. Hence, the metric there is
flat to leading order, so the leading terms in \eomsem\ and
\viraco\ should be equal to the flat space ones \eomfive\ and
\vira. Let us see how this works.

For the equations of motion \eomsem\ we need the variation of the
metric \offmetric. If we evaluate \eomsem\ around the location of
the bent NS' at $z\sim a_1$, and then plug the boundary condition
$w=\tan\theta\,v$, the $v$ and $w$ equations of motion become
equal and read
 \eqn\veom{
 \eqalign{
 \left[1-{1\over3}\sum{1\over
 r^3}\right]\partial_z\bar\partial_{\bar z}\bar
 v+{1\over2}s\bar\partial_{\bar z}\bar s\partial_z \bar v\sum{1\over
 r^5}=0 \ .
 }}
where we introduced the short-hand notation
 $$
 \sum_{i=1}^{k-1}{c\over\vert r-r_i\vert^\alpha_\perp}\equiv\sum{1\over r^\alpha}.
 $$
For $s$ we get an analogous expression. On the other hand,
evaluating the Virasoro condition \viraco\ around $z\sim a_1$ and
using the metric \offmetric, once again we can plug directly the
boundary condition $w=\tan\theta\,v$ and we get
 \eqn\viraci{
 \left[\left(1-{1\over3}\sum{1\over
 r^3}\right)(1+\tan^2\theta)+\sin^2\theta\,\sum{1\over
 r^3}\right]\partial_z v \partial_z \bar
 v+\left[1+{2\over3}\sum{1\over r^3}\right]\partial_z s\partial_z
 \bar s =0 \ ,
 }
which is now an equation for the asymptotic behavior of $v,s$
around $z\sim a_1$.

Finally, one has to solve \veom\ and \viraci\ for the embedding
functions $v,w,s$ around the pole at $z\sim a_1$. Note that the
leading terms in the radius expansion are just the flat space
equations \eomfive\ and \vira. But we have already seen that there
is no non-holomorphic solution to these equations. Hence, we
conclude that the M5-brane framework does not exhibit the
metastable gauge theory vacuum, even when including the
backreaction of the spectator components of the fivebrane.

\newsec{M5-brane non-supersymmetric vacua: soft
 breaking}

In this Section we will consider M5-branes wrapping minimal volume
non-holomorphic curves with non-holomorphic boundary conditions.
We will later interpret this type of supersymmetry breaking as the
analog of soft supersymmetry breaking in gauge theory. While in
the case of a holomorphic rotation $w=mv$ only the monopole and
dyon points are lifted to ${\cal N}=1$ vacua, we will see that
with the non-holomorphic rotation $w=m(v+\bar v)$ any point on the
${\cal N}=2$ moduli space is lifted to a non-supersymmetric
vacuum.

\subsec{A non-holomorphic torus}

We have seen that the M5-brane does not exhibit the analog of
metastable gauge theory vacua. In particular, it is not possible
to realize the holomorphic boundary conditions \massns\ with a
minimal volume non-holomorphic curve. In the following we will
look for a non-holomorphic M5-brane configuration which is as
close as possible to the one in \massns. We will take the boundary
conditions to be non-holomorphic but still with a {\it linear}
relation between $w$ and $v$. As we will discuss, this may
describe the analog of soft supersymmetry breaking in the gauge
theory.

We look for a minimal area curve that satisfies the following
conditions:

\noindent $i)$ It is  a genus one curve with {\it non-holomorphic}
embeddings in the target space coordinates $(w,v,s)$. The
embeddings have to be harmonic functions of $z$ as in \harm.

\noindent $ii)$ We consider a rotation with a mass parameter $m$,
such that, when we take $m$ to zero, we recover the ${\cal N}=2$
curve \embetwo. Thus, our exact solution can be considered for
small $m$ as a perturbation of the ${\cal N}=2$ theory.

\noindent $iii)$ We require that at infinity, at first order in
$m$ and in $g_s$, the rotation is of the form
 \eqn\rotation{
 w\sim(v+\bar v) \ .
 }
Note, that if $w$ is proportional to a higher power of $v$, it is
hard to solve the Virasoro constraint, since the higher order
poles coming from the $w$ contribution cannot be cancelled.

\noindent $iv)$ We fix the periods and residues of $ds$ as in
\resin, \aper, \bconst, \bper.

The requirements that $w$ diverges linearly at the NS and the NS'
branes means that it can only depend on $F^{(1)}(z-a_i)$ and its
complex conjugate, but not on higher derivatives $F^{(k>1)}$,
since $F^{(n)}(z-a_i)$ has an n-th order pole at $z=a_i$.
Moreover, requiring that it is harmonic and elliptic fixes its
form uniquely to
 \eqn\wnew{
 w=mg_sA\left(F^{(1)}(z-a_1)+F^{(1)}(z-a_2)+\overline{F^{(1)}(z-a_1)}+\overline{F^{(1)}(z-a_2)}
 \right) \ ,
 }
where the factor of $g_s$ will be necessary to satisfy the
Virasoro constraint \vira. The embedding function $v$ can be
parameterized as
 \eqn\vlast{\eqalign{v=&A\left(F^{(1)}(z-a_1)-F^{(1)}(z-a_2)-i\pi\right)+\half
u_1-Ag_s^2
 m^2\left(\overline{F^{(1)}(z-a_1)}-\overline{F^{(1)}(z-a_2)}
 \right) \cr
 &+m^2g_s^2\rho\left(F^{(1)}(z-a_2)+\overline{F^{(1)}(z-a_2)}\right) \ ,
 }}
where the first term is the SW solution \embetwo\ and we fix the
overall scale $A(\tau)$ by consistency with the $m=0$ limit.
Finally, the embedding function $s(z,\bar z)$ acquires an
anti-holomorphic part as well
 \eqn\semb{\eqalign{
 s=&2(F(z-a_1)-F(z-a_2)-i\pi z)\cr
 &+m^2\gamma\left(F(z-a_1)-F(z-a_2)+\overline{F(z-a_1)}-\overline{F(z-a_2)}-i\pi(z-\bar
z)\right) \ ,
 }}
and note in particular that its holomorphic and anti-holomorphic
parts are
 \eqn\ssw{\eqalign{
 s_H=\left(1+{m^2\gamma\over2}\right) s_{SW}(z) \ ,\qquad
 s_A={m^2\gamma\over2}s_{SW}(z)\ .
 }}
One can easily check that this satisfies our period conditions
\aper\ and \bconst\ that fix the distance between the two marked
points on the torus, namely $a\equiv a_2-a_1 =-\tau/2$, as in the
${\cal N}=2$ case (see appendix B). The last boundary condition
\bper\ fixes the dependence of the period $\tau$ of the torus
boundary data at the cutoff scale
 \eqn\running{\eqalign{
 2\pi i
 \alpha(v_0)=&-2\left(\ln\left({v_0\over
 A}\right)^2+2F(\tau/2)-2\ln\theta'(\tilde\tau)-i\pi\tau/2\right)
 \cr
 &-m^2\gamma\left(\ln\left({v_0\over
 A}\right)^2+2F(\tau/2)-2\ln\theta'(\tilde\tau)-i\pi\tau/2+c.c.\right)
 }
 }
Since we have fixed $A(\tau)$ by the ${\cal N}=2$ limit, \ssw\
takes a very simple form
 \eqn\ssimple{
 2\pi i\alpha(v_0)=-2\ln\left({v_0\over\Lambda}\right)^2-m^2\gamma\left\vert{v_0\over\Lambda}\right\vert^4 \ .
 }
Here and in the following we still will denote by $\Lambda$ the
scale of the unperturbed ${\cal N}=2$ theory.

The parameters $\rho,\gamma$ in \vlast\ and \semb\ are fixed by
the Virasoro condition \vira. As anticipated above, the way to
satisfy this constraint is the following. The functions appearing
in the constraint are all elliptic functions. Hence, we just need
to expand them around a pole, say $z=a_1$, and require that the
coefficients of the poles of different degrees and the constant
term in the expression vanish. This fixes the various coefficients
in the embedding functions. The fourth order pole has already been
cancelled by the $m^2g_s^2$ term in \vlast, so we are left with a
double pole, a single pole and a constant, each of which must
vanish separately. The details of this computation are given in
appendix C. The result is that $\rho,\gamma$ depend on
$\tau,m,g_s$ as follows
 \eqn\coeffs{\eqalign{
 \gamma=&-{1\over m^2}+{1\over m^2}\left[
 {\alpha+(1-m^2g_s^2)A^2\beta^{1\over2}\over 8g_s^2(\wp+2\eta_1)}\right]^{1\over2}
 \ , \cr
 \rho=&{1\over m^2g_s^2-1}\left(4A+{2\gamma+m\gamma^2\over Am(\wp+2\eta_1)}\right) \ ,
 }}
where $\alpha=\alpha(\tau,g_s,m)$, $\beta=\beta(\tau,g_s,m)$, the
Weierstrass function $\wp(\tau/2)$ and $\eta_1(\tau)$ are certain
elliptic functions defined in Appendix C. We can look at their
leading order expansion as we take $m\to0$ to check that we get
back the ${\cal N}=2$ solution of \embetwo\ in this limit. Both
$\gamma$ and $\rho$ are indeed finite in this limit. In Fig.5 and
6 we plot the two coefficients as functions of $\tau$ imaginary.
\ifig\gammatt{Plot of the coefficient $\gamma(\tau)$ in \coeffs\
for imaginary values of $\tau$ and small $m$. It is monotonic for
any $m$ and intercepts $\gamma(\tau=0)=\Lambda^2/3$.}
{\epsfxsize3in\epsfbox{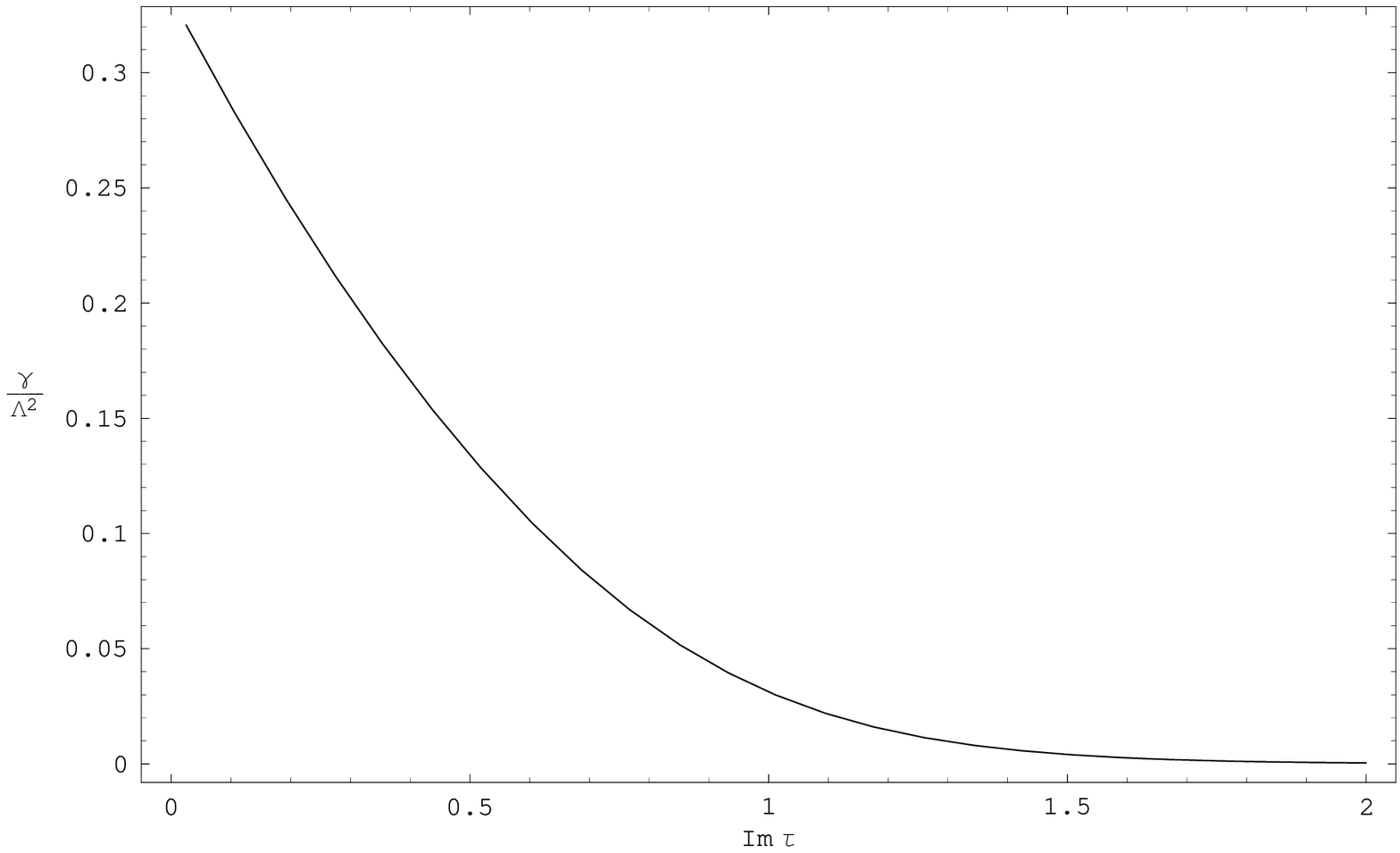}} \ifig\rhott{Plot of the
coefficient $\rho(\tau)$ in \coeffs\ for imaginary values of
$\tau$. It always vanishes at the origin. It also vanishes at a
finite value of ${\rm Im}\,\tau$ for small $m$, in the left plot.
When $m$ increases the local maximum disappears and $\rho$ becomes
monotonic and negative, on the right plot for $m=1$.}
{\epsfxsize2.5in\epsfbox{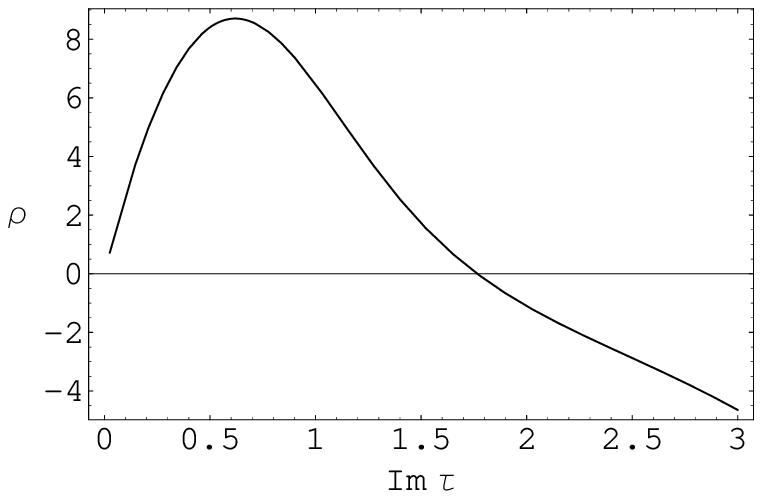}\epsfxsize2.5in\epsfbox{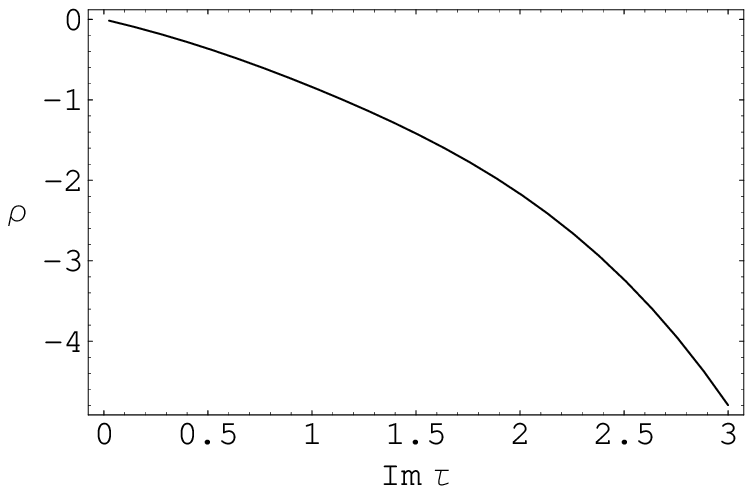}}

Let us look at the boundary conditions at the NS and NS' brane,
coming from our exact solutions. At first order in $m g_s$ we have
 \eqn\bending{
 w=\pm mg_s(v+\bar v)\ .
 }
Hence the embedding $w$ grows linearly in $v$ at infinity, as
expected for a mass term in the gauge theory, while the $s$
embedding bends logarithmically in $v$, as seen in \ssimple, as
appropriate for a running coupling. Note that, since
$\gamma(\tau)$ and $\rho(\tau)$ have a smooth $\tau\to0$ limit, we
can rotate in a non-holomorphic way the dyon and monopole points
as well, where the ${\cal N}=2$ torus degenerates to a sphere.

Let us comment on the behavior of our solution as we increase the
mass parameter $m$, that controls the rotation of the NS' brane
away from the ${\cal N}=2$ point. The coefficient $\gamma(\tau)$
is a monotonically decreasing function of $\tau$ and it drops to
zero for large values of $\tau$. There is a very interesting
behavior of $\rho(\tau,m)$ as we vary $m$. When $m$ is very small,
$\rho$ starts at zero, where the torus degenerates to a sphere,
and has a local positive maximum, then it vanishes again at a
finite value of $\tau=\hat\tau$. The points for which $\rho$
vanishes have actually the same bending at infinity in the $w$ and
$v$ directions
 \eqn\exact{
 w=\pm {mg_s(v+\bar v)\over 1-m^2g_s^2} \ .
 }
However, their logarithmic bending in the $s$ direction depends on
the coefficient $\gamma(\tau)$, which is monotonic. In the case in
which $x_6$ is non-compact, the sphere at $\tau=0$ and the torus
at $\tau=\hat\tau$ have different subleading logarithmic bending.
Hence they have the same boundary conditions in the directions $w$
and $v$ but not in the direction $s$.

\subsec{Scalar Potential}

We have found that any point $\tau$ in the ${\cal N}=2$ moduli
space can be lifted to a non-holomorphic curve. Every $\tau$ gives
a different choice of boundary conditions, so we have a one
parameter family of non-supersymmetric solutions to the
supergravity equations. We can evaluate the M5-brane action on our
solution, that is the volume of the curve, which corresponds to
computing the scalar potential depending on the parameter $\tau$.
The action is infinite and needs to be regularized. The divergent
part comes from the fact that the curve is non-compact, namely the
embedding functions $(w,v,s)$ have poles. A way to regularize the
action is to isolate in \areael\ the term that, upon integration,
is proportional to the spacetime Kahler form \deBoerBY. What is
left is the integral of $G_{i\bar j}\partial_{\bar z}X^i\partial_z
X^{\bar j}$, which vanishes for holomorphic curves. This
regularization is appropriate when the boundary conditions are
holomorphic. In fact, when the full curve is holomorphic, the
action is zero, as expected for the energy of a supersymmetric
vacuum. When the curve is non-holomorphic but the boundary
conditions are still holomorphic, the action is finite and
positive, and corresponds to the fact that a dynamical
supersymmetry breaking vacuum has positive energy. In our case,
however, the boundary conditions are non-holomorphic and this
regularization, although possible, will give an infinite result
anyway. Hence, we regularize the action by introducing a cutoff in
the spacetime variable $v_0$. At this scale we specified the
boundary conditions for the B period of $ds$, the running gauge
theory coupling at the cutoff scale. The action of the fivebrane
in the eleven dimensional supergravity approximation is \areac,
that we can rewrite in our simple case of flat metric as
 \eqn\fullpot{
 A={1\over g_s^2}\int_\Sigma \left(dv\wedge*d\bar v+
 dw\wedge*d\bar w+g_s^2 ds\wedge*d\bar s\right) \ .
 }
The contribution to the potential coming from $w$ and $v$
coordinates, which is usually subleading, in this case must be
taken into account, since all of the terms are of the same order.
It is straightforward to evaluate the integral \fullpot\ by using
the Riemann bilinear relations, properly regularized to take into
account the divergences. The contribution from the $A$ and $B$
cycles vanish and we are left with just the integral around the
marked points. The embedding functions $w$ and $v$ give a
quadratic divergence in the cutoff $v_0$, whose coefficients
depend on the modulus $\tau$
 \eqn\dvdv{\eqalign{
 \int_\Sigma dv\wedge*d\bar v=&v_0^2\left(1+\left(1-{m^2g_s^2\rho\over A}\right)^2\right)+
 \bar v_0^2\left(m^4g_s^4+m^4g_s^4\left(1-{\rho\over A}\right)^2\right) \
 , \cr
 \int_\Sigma dw\wedge*d\bar w=&2m^2g_s^2(v_0^2+\bar v_0^2)\
  \ ,
 }}
where the term that do not depend on $m$ are the ${\cal N}=2$
contributions and $\rho(\tau)$ and $A(\tau)$ are in \coeffs\ and
in the Appendix B. The contribution to the action \fullpot\ by the
$s$ embedding function is given by
 \eqn\potentorus{
 \int_\Sigma ds\wedge*d\bar s=(2+m^2\gamma)^2\ln\left({v_0\over
 \Lambda} \right)^2+m^4\gamma^2\ln{\left({\bar v_0\over
 \bar \Lambda} \right)}^2 \ ,
 }
The first correction comes in at order $\tilde m^2=m^2g_s^2$
 \eqn\potenm{
 V\approx V_{{\cal N}=2}+2m^2g_s^2\left(v_0^2+\bar v_0^2-{\rho\over
 A}v_0^2+2\gamma\ln\left({v_0\over
 \Lambda} \right)^2\right) \ ,
 }
where $\rho,\gamma$ and $A$ are functions of the modulus $\tau$.
The ${\cal N}=2$ potential $V_{{\cal N}=2}$ is just a constant
term and do not depend on the modulus $\tau$. Since the divergent
part depends on $\tau$, the various non-holomorphic curves differ
by an infinite amount of energy, so we cannot compare them. Note
that the quadratically divergent part of the potential depends on
$\tau$ through the combination $\rho/A$. In Fig.6 we have shown
that, for small $m$, there are two different values of $\tau$ such
that $\rho$ vanishes. Hence, for these two different curves (a
sphere and a torus) the leading divergence in the energy is the
same, however the logarithmic bending is still different, so they
do not represent a metastable pair of vacua.

\newsec{Soft terms in the gauge theory limit}

In this Section we want to interpret the non-holomorphic torus we
have found in terms of the gauge theory. Usually, the boundary
conditions at infinity correspond in the gauge theory to the
choice of the classical ${\cal N}=1$ superpotential. Our boundary
conditions at infinity \bending\ correspond to a non-holomorphic
quantity in the gauge theory, so that supersymmetry is explicitly
broken by a soft term (a non-supersymmetric relevant deformation).
To see this, it is more convenient to shift the embedding
coordinate $w$ so that the asymptotics at the NS brane, located at
$z\sim a_2$, is the more familiar $w\sim 0$. Then, the asymptotics
at the NS' brane at $z\sim a_1$ represents the rotation of the NS'
brane with respect to the NS brane. As usual, we can identify the
embedding coordinate $v=x_4+ix_5$ with the eigenvalues of the
adjoint operator $\Phi$ in the gauge theory, by matching their
$U(1)_R$ charge. To leading order in the string coupling, the NS'
brane at infinity is rotated with respect to the NS brane by the
amount
 \eqn\rota{
 w= 2 mg_s(v+\bar v)+{\cal O}(g_s^2) \ ,
 }
The gauge theory limit is given as usual by taking $g_s,l_s,\Delta
L\to0$ while keeping the Yang-Mills coupling
$g_{YM}^2=g_sl_s/\Delta L$ fixed, where $\Delta L$ is the distance
between the NS and NS' fivebranes. The mass $\tilde m$ in the
gauge theory is related to the string quantities by
 \eqn\mass{\tilde m={g_sm\over l_s} \ .
 }

The boundary conditions \rota\ resemble a gauge theory mass term
for a real component of a chiral superfield. In our ${\cal N}=2$
gauge theory we only have $\Phi$, which transforms in the adjoint
representation of the gauge group $U(2)$. Since we do not want to
break explicitly gauge invariance, the only field whose real part
can get a mass term is the $U(1)$ part of the adjoint, that we
denoted $u_1=\Tr \Phi$. The deformation \rota\ corresponds to a
soft supersymmetry breaking mass term $\tilde m$ for the scalar
component of ${\rm Re} (u_1)$
 \eqn\softla{
 {\cal L}_{soft}={1\over4} \tilde m^2  (u_1+u_1^\dagger)^2\ .
 }
It is easy to see that such a soft mass term can be obtained by
the following term in the lagrangian
 \eqn\supersoft{
 {\cal L}_{soft}=\int d^4\theta Z(X,X^\dagger)u_1^\dagger u_1+
 \int d^2\theta {{M}}\,u_1^2+{\rm h.c.} \ ,
 }
by promoting the wavefunction renormalization $Z(X,X^\dagger)$ and
the bare mass $M$ to spurions with non-zero F and D components, in
the case of the real superfield $Z$, or only F components, in the
case of the chiral superfield $M$ \GiudiceNI. This is what happens
when integrating out a massive messenger sector, that couples the
visible sector, represented by our ${\cal N}=2$ theory, to a
hidden sector $X$, that breaks supersymmetry spontaneously by
acquiring an F-term $\langle X\rangle=\Lambda_{susy}+\theta^2
F_X$. In the simplest case we can take $Z=X^\dagger
X/\Lambda^2_{susy}$. By appropriately choosing the expectation
values of $M$ and $X$, one can then easily reproduce the soft mass
term \softla.

Our non-supersymmetric brane configuration therefore realizes
${\cal N}=2$ gauge theory in which we first softly break to ${\cal
N}=1$ by a superpotential term and, in a second step, we break to
${\cal N}=0$ by coupling it to a hidden sector through a massive
messenger interaction.

\vskip0.2cm \centerline{\bf Acknowledgements}

We would like to thank Ben Burrington, Sunny Itzhaki and Stefan
Theisen for discussions. L.M. would like to thank Joe Marsano and
Masaki Shigemori for very useful discussions and correspondence
and the organizers of the Simons Workshop 2007 at Stony Brook for
the kind hospitality, where part of this work has been done.

\appendix{A}{Elliptic functions}

The main object for the construction of the elliptic (i.e. doubly
periodic) functions
 \eqn\build{
 F(z)=\ln \theta\left(\pi(z-\tilde\tau)\right) \ ,
 }
where $\theta(z)\equiv \theta_3(z,q)$ is the standard Jacobi theta
function that has a zero at $-\pi\tilde \tau$ where
$\tilde\tau=\half(\tau+1)$.\foot{We follow here the convenient
notations and conventions in \MarsanoFE\JanikHK, to which we refer
the interested reader.} Hence, $F(z)\sim\ln z$ at $z\sim0$. The
$n$-th derivative of $F(z)$ has an $n$-th order pole at $z=0$, so
let us introduce the notations
 \eqn\notat{
 F_i^{(n)}=\partial_z^nF(z-a_i) \ ,
 }
 and the $F_i^{(n)}$ are elliptic for $n>1$ and have the following
monodromies for $n=0,1$
 \eqn\monodr{\eqalign{
F_i(z+1)&=F_i(z),\cr F_i(z+\tau)&=F_i(z)+i\pi - 2\pi i(z-a_i),\cr
F_i^{(1)}(z+1)&=F_i^{(1)}(z),\cr
 F_i^{(1)}(z+\tau)&=F_i^{(1)}(z)-2\pi i.
 }}
The $F_i^{(n)}$ also have nice properties under $z\rightarrow -z$
 \eqn\reflex{\eqalign{
F(-z) &= F(z)-2\pi i z+i\pi,\cr F^{(1)}(-z)&=-F(z)+2\pi i,\cr
F^{(n)}(-z) &= (-1)^nF(z)\qquad n>1 \ ,
 }}
and, most importantly, their half period value is zero for the odd
derivatives
 \eqn\halfper{
 F^{(2n+1)}(\tau/2)=0 \ .
 }
The asymptotic expansion of $F^{(1)}(z)$ around the origin is
 \eqn\asymori{
 F^{(1)}(z)={1\over z}+i\pi-2\eta_1z-{g_2z^3\over 60}+{\cal
 O}(z^5) \ .
 }

The basic Weierstrass elliptic function is defined as
 \eqn\weierser{
 \wp(z)={1\over z^2}+\sum_{m,n=-\infty}^\infty\left({1\over
 (z-(m+\tau n))^2}-{1\over (m+n\tau)^2}\right), \qquad (m,n)\neq
 (0,0)
 }
and the Weierstrass zeta and sigma functions are defined by
$\wp(z)=-\partial_z\zeta(z)$ and
$\zeta(z)=\partial_z\ln\sigma(z)$. The Weierstrass functions are
related to $F(z)$ as follows
 \eqn\fweier{\eqalign{
 F(z)=&\ln[\sigma(z)\theta'(\tilde\tau)]-\eta_1z^2+i\pi z \ ,\cr
 F^{(1)}(z)=&\zeta(z)-2\eta_1z+i \pi \ ,\cr
 F^{(2)}(z)=&-\wp(z)-2\eta_1 \ .
 }}
The Weierstrass function satisfies the differential equation
 \eqn\differen{
 \partial_z\wp(z)^2=4\wp(z)^3-g_2\wp(z)-g_3 \ .
 }
It proves useful to rewrite some of these objects in terms of
Jacobi theta functions
 \eqn\jacobiwe{\eqalign{
 \wp(\tau/2)=&-{\pi^2\over3}\left[\theta_2^4(0)+\theta_3^4(0)\right]
 \ ,\cr
 \eta_1(\tau)=&\zeta(\half)=-{\pi^2\over6}{\theta_1'''(0)\over
 \theta_1'(0)} \ , \cr
 g_2(\tau)=&{2\pi^4\over3}\left[\theta_2^8(0)+\theta_3^8(0)+\theta_4^8(0)\right]
 \ ,
 }}

\appendix{B}{The parametric ${\cal N}=2$ curve}

The brane configuration
 \eqn\branetwo{
\matrix{    & x_0     & x_1    & x_2 & x_3 & v & x_6 & x_7 & w\cr
          NS & \bullet & \bullet &\bullet &\bullet  &\bullet
          &\times &\times &\times \cr
          NS\,'& \bullet & \bullet &\bullet &\bullet  & \bullet
          &\times &\times &\times \cr
          D4& \bullet & \bullet &\bullet &\bullet  & \times
          &\bullet &\times &\times \cr
           }
 }
describes ${\cal N}=2$ gauge theory with $U(N)$ gauge group. Its
lift to M theory\foot{For related work, see \HoweEU.} \WittenSC\
is an M5-brane wrapping the holomorphic curve of genus $N-1$
 \eqn\curvetwo{
 \Sigma_c:\qquad \left\{\matrix{&t^2-2tP_N(v,u_r)+4\Lambda^{2N}=0 \ ,\cr
 &w=0\ ,    }\right.
 }
plus a bunch of disconnected complex lines, in the case we have
also flat spectator NS' branes. Let us consider the case in which
the gauge group is $U(2)$, i.e. we have two D4-branes. In this
case the curve is a torus. We would like to give a parametric
description of \curvetwo\ in the $z$ coordinate, \foot{We use
similar techniques to \JanikHK\MarsanoFE, who studied the ${\cal
N}=1$ case.} by using elliptic functions. The embedding functions
$s$ and $v$ are holomorphic and completely fixed by their periods
and their asymptotic boundary conditions to
 \eqn\embet{
 \eqalign{
 s_{SW}(z)=&2\left(F(z-a_1)-F(z-a_2)\right)-2\pi i z +s_0\ ,\cr
 v_{SW}(z)=&A\left(F^{(1)}(z-a_1)-F^{(1)}(z-a_2)-i\pi\right)+\half
 u_1 \ ,
 }}
and $w(z)=0$, while the B-period constraint \bconre\ fixes
$a=-\tau/2$. We would like to find the map between the parametric
quantities $\tau,A,s_0$ in \embet\ and the physical quantities
$u_1,u_2,\Lambda$ in \curvetwo. Noting that $t=e^{-s}$, by
plugging \embet\ into \curvetwo\ with characteristic polynomial
$P_2(v)=v^2-u_1v-u_2+\half u_1^2$ we eventually find the exact map
between the parametric and the physical quantities
 \eqn\utau{\eqalign{
 A^2(\tau)=&-\Lambda^2\left(12\wp(\tau/2)^2-g_2(\tau)\right)^{-\half} \
 , \cr
 u_2(\tau)=&-3\wp(\tau/2)A^2(\tau)+{u_1^2\over4} \ . \cr
 }}
The parametrization of the moduli space using $\tau$ is actually a
multiple covering. A part from the obvious symmetry
$u(\tau+2)=u(\tau)$, there are also the reflection symmetries
\BonelliRY\ $u(\tau+1)=-u(\tau)$ and
$u(-\bar\tau)=\overline{u(\tau)}$. The three punctures are at
$u(\tau=0)=\Lambda^2$, $u(\tau=1)=-\Lambda^2$ and
$u(\tau=\infty)=\infty$.

\appendix{C}{Virasoro condition for the non-holomorphic torus}
In this Appendix we give some details about the computation of the
exact non-holomorphic solution in \wnew, \vlast\ and \semb.

Let us discuss the conditions on the harmonic embedding coordinate
$s$. The most general elliptic and harmonic function satisfying
the period conditions \aper, \bconst\ is
 \eqn\semb{\eqalign{
 s=&(2+m^2\gamma)(F_1-F_2)-i\pi(2-m^2\delta)z\cr
 &+m^2\gamma(\overline{F_1}-\overline{F_2})-i\pi m^2\delta\bar
 z \ ,
 }}
where $\gamma,\delta,A$ are constant coefficients to be fixed. The
 condition \bconst\ gives
 \eqn\bcons{
-4a-2\tau+m^2\delta(\tau-\bar\tau)-2m^2\gamma(a-\bar a)=0 \ ,
 }
which fixes the location of the two marked points on the torus as
in \bconre\ and is satisfied by $a=-\tau/2$ as in the ${\cal N}=2$
case, with $\delta=-\gamma$. The B-period in \bconst\ then fixes
the dependence of $\tau$ on the running coupling of the gauge
theory at the cutoff scale. The parameters $\rho,\gamma,\delta$ in
our ansatz \wnew, \vlast\ and \semb\ are going to be fixed by
solving the Virasoro condition
 \eqn\viras{
 g_s^2\partial s\partial \bar s+\partial v\partial \bar v+\partial
 w\partial\bar w=0 \ .
 }
As explained in the main text, we just need to expand \viras\
around a pole, say $z=a_1$, and impose that the coefficients of
the poles of different degrees and the constant term in the
expression separately all vanish. The quartic pole cancels
automatically, while we have again three complex equations coming
from the double pole, the simple pole and the constant term. At
the special point $a=\tau/2$, solution of \bcons, we have
$\delta=-\gamma$ and the odd derivatives $F^{(2n+1)}(\tau/2)=0$.
The equations simplify and allow to solve for the real and
imaginary parts of the coefficients in the game, namely
$\gamma,\rho$. We want the solutions to satisfy the requirement
that, in the limit $m\to0$, we recover the ${\cal N}=2$ solution
\embetwo, so $\gamma,\rho$ must be finite. It turns out that there
is a unique solution satisfying these requirements.

The exact solution is
 \eqn\second{\eqalign{
 \gamma&=-{1\over m^2}+{1\over m^2}
 \left({\alpha(\tau,m,g_s)+(1-m^2g_s^2)A^2\beta(\tau,m,g_s)^{1\over2}\over8g_s^2(\wp+2\eta_1) }\right)^{1\over2} ,\cr
 \rho&={1\over (m^2g_s^2-1)}\left(4A+{\gamma(2+m\gamma)\over Am(\wp+2\eta_1)}\right)\ ,
 }}
where we introduced the elliptic functions
 \eqn\elli{\eqalign{
\alpha(\tau,m,g_s)=&8g_s^2(\wp+2\eta_1)\left(1-4m^2A^2(\wp+2\eta_1)\right)-(1-m^2g_s^2)^2A^2h(\tau)\
,\cr
 \beta(\tau,m,g_s)=&(1+m^2g_s^2)^2h(\tau)^2-4m^2g_s^2(h(\tau)-8(\wp+2\eta_1)^2)^2
 \ ,\cr
 h(\tau)=&24\wp(\tau/2)^2+24\wp(\tau/2)\eta_1(\tau)-g_2(\tau) \ .
 }}

The solution can be expanded to first order in the mass
 \eqn\seconda{\eqalign{
 \gamma&\sim-8A^2{(\wp+2\eta_1)^3\over h(\tau)}+{\cal O}(m^3) ,\cr
 \rho\,&\sim -4A\left(1-4{(\wp+2\eta_1)^2\over h(\tau)}\right)+ {\cal O}(m^3)\ ,\cr
 }}
where $\wp$ is the Weierstrass $\wp$-function evaluated at
$\tau/2$ and $\eta_1$ and $g_2$ are and some standard coefficients

\listrefs

\bye